\documentclass[twocolumn]{aastex62}

\newcommand{\kms}{\ensuremath{\mbox{km s}^{-1}}}
\newcommand{\rphk}{\ensuremath{R'_{\rm HK}}}
\newcommand{\lrphk}{\ensuremath{\log{\rphk}}}

\shorttitle{A Maunder Minimum Candidate}
\shortauthors{Shah et al.}

\begin{document}

\title{HD 4915: A Maunder Minimum Candidate}

\author{Shivani P. Shah}
\affiliation{Center for Exoplanets and Habitable Worlds, Department of Astronomy \& Astrophysics, The Pennsylvania State University, 525 Davey Laboratory, University Park, PA 16802, USA}
\author{Jason T.\ Wright}
\affiliation{Center for Exoplanets and Habitable Worlds, Department of Astronomy \& Astrophysics, The Pennsylvania State University, 525 Davey Laboratory, University Park, PA 16802, USA}
\author{Howard Isaacson}
\affiliation{Department of Astronomy, 601 Campbell Hall, University of California, Berkeley, CA 94720-3411, USA}
\author{Andrew Howard}
\affiliation{California Institute of Technology, 1200 E California Boulevard, Pasadena, CA, 91125, USA}
\author{Jason L.\ Curtis}
\altaffiliation{NSF Astronomy \& Astrophysics Postdoctoral Fellow}
\affiliation{Department of Astronomy, Columbia University, 550 West 120th Street, New York, NY 10027, USA}

\begin{abstract}
We study the magnetic activity cycle of HD 4915 using the \ion{Ca}{2} H \& K emission line strengths measured by Keck I/HIRES spectrograph. The star has been observed as a part of California Planet Search Program from 2006 to present. We note decreasing amplitude in the magnetic activity cycle, a pattern suggesting the star's entry into a Magnetic Grand Minimum (MGM) state, reminiscent of the Sun's Maunder and Dalton Minima. We recommend further monitoring of the star to confirm the grand minimum nature of the dynamo, which would provide insight into the state of the Sun's chromosphere and the global magnetic field during its grand minima. We also recommend continued observations of H \& K emission lines, and ground or space based photometric observations to estimate the sunspot coverage.
\end{abstract}


\section{INTRODUCTION}
It has been almost 175 years since Heinrich Schwabe discovered the 11 year sunspot cycle in 1843. With the development of solar dynamo theory as the mechanism responsible for creating the Sun's magnetic field, the sunspot cycle was realized to be a manifestation of the Sun's magnetic cycle (see \cite{dynamo} for a review). About 140 years after the discovery of sunspot cycle, \cite{Eddy} compiled strong historical evidence spanning centuries to show that the "prolonged sunspot minimum" period from 1645 to 1715 as speculated by Gustav Sp\"{o}rer in 1889 and E. W Maunder in 1894 was real. Eddy called this period the Maunder Minimum. Two other such periods, although less dramatic, were identified by the community \citep[e.g.,][]{Siscoe}, the Dalton (1796-1820) and Sp\"{o}rer Minima (1460-1550).\footnote{The origin of the term ``Dalton Minimum'' is unclear, but the first usage we are aware of is \citet[][p.59]{1982proceedings}. On the other hand, Sp\"{o}rer Minimum was first used in \cite{Eddy}} Such periods are now more generally termed Magnetic Grand Minima (MGM) \citep[e.g.,][]{Saar} and are characterized by repressed magnetic activity that lasts longer than at least the typical cycle of the star.
\par 
Learning more about MGM periods has been of interest because it addresses one of the many elusive characteristics of the solar dynamo, thereby contributing to our understanding of the Sun's magnetic field, especially its evolution in the past and future. To maximize this learning, several surveys of Sun-like stars have been conducted, with the goal of collecting a large sample of stars in MGM states. 
\par
Today, the standard metric for sunspot coverage is Group Sunspot Number (GSN), which uses the number of groups of sunspots to provide a more self-consistent and less noisy dataset than the previous calibrations \citep{Hoyt}. Because it is  challenging or impossible to observe sunspots on Sun-like stars on decade timescales, it is necessary to use a proxy for global magnetic field strength instead, such as chromospheric line emission. In this regard, the most common proxy has been the emission cores of the Ca II H\&K lines at 3968 \AA\ and 3934 \AA, respectively. 
\par 
A drawback of using this proxy is the absence of chromospheric line emission data from the MGM periods of the Sun, specifically the Maunder Minimum, preventing an absolute comparison in the activity of the Sun and other stars. This compels the use of sunspots instead; however, it is important to note that although, the sunspots have shown an almost flat activity during the Maunder Minimum period, other solar proxies, including the cosmogenic isotopes \citep{Kocharov}, have in fact, indicated a weekly cycling activity.  
\par
\subsection{Searching for MGM in other stars}
The oldest survey to track the Ca II H\&K line emission was the Mount Wilson Observatory HK Project that examined bright nearby stars for photospheric and chromospheric variability for decades \citep{Wilson1968,S-Index,Sally95}. With the extensive data available through this survey, the search for MGM candidates followed; this search has now spanned decades, using data from some of the most extensive surveys. The biggest challenge, however, has been to set down the criteria for a good MGM candidate based on observations of magnetic activity levels, metallicity, age, stellar type, gravity, and the relation between these properties (e.g., \cite{Sally&jastrow}, \cite{Henry}, \cite{Saar}). 

One of the earliest attempts to set down the criteria was made by \cite{Sally&jastrow}, as they investigated the magnetic activity distribution of 74 solar-type stars from the Mount Wilson HK Project using the Mount Wilson S-Index measurement. Their selection criteria for Sun-like stars were: (1) $B-V$ in the range of 0.60-0.76 (the Sun has $B-V=0.65$) and (2) an age similar to the Sun.

\cite{Sally&jastrow} found that the frequency distribution of these 74 stars had a distinct bimodal feature with a broad peak centered at $S \approx 0.17$ and a narrow peak centered at \textit{S} $\approx$ 0.145, lower than the S-Index of Sun's typical minimum, \textit{S} = 0.164 (Figure 2 in \citep{Sally&jastrow}). They suggested that the broad peak represented stars undergoing cyclic variations like the Sun and the narrow peak represented stars in Maunder minimum (MM). They noted that 4 of these MM candidates showed "flat" activity time series concluding that 30\% of Sun-like stars are in states analogous to Maunder Minimum of the Sun.
\par 
Based on this, \cite{Sally&jastrow} proposed that future MM candidates could be identified by a combination of "flat" activity and an average $S < 0.15$ i.e., should contribute to the narrow peak. 
\par 
\cite{Henry} conducted a similar investigation, but on a sample of stars surveyed as a part of the Project Phoenix Survey \citep{Phoenix} and found a similar bimodal distribution. \cite{Henry} further formed a subclass of very inactive stars with $\lrphk < -5.1$, corresponding to $S \approx 0.15$.
\par
However, using accurate parallaxes from the \textit{Hipparcos} Catalogue \citep{hipparcos}, \cite{Wright} showed that almost all MM candidates in the Mount Wilson and Project Phoenix surveys were actually of luminosity class IV/V or IV (i.e., either slightly evolved or subgiants). As had been shown by \cite{Nascimento}, evolved stars generally have lower S-Indices than dwarf stars of the same color, and at any rate evolved stars are, by definition, not Sun-like, and so do not belong in these samples. Because the effects of modest evolution are only weakly discernable in stellar spectra (apparent only as subtle changes in gravity-sensitive lines), surveys of Sun-like stars conducted prior to \textit{Hipparcos} suffered significant contamination from subgiants due to the Malmquist bias. 
\par
As a part of another effort, \cite{Giampapa} found in their survey of M67 cluster that 17\% of Sun-like stars exhibited noticeably smaller magnetic activity levels than the typical levels of the Sun and categorized them to be MM candidates. \cite{Giampapa} addressed this phenomenon qualitatively and suggested that Maunder Minimum like periods were low-amplitude extension of the solar dynamo rather than a state in a separate modal frequency as was indicated by \cite{Sally&jastrow} and \cite{Henry}. 
\par
However, \cite{Curtis} demonstrated that the low activity levels were an effect of the contamination of the H\&K lines due to calcium absorption in the ISM. On correcting the HK line strengths for likely levels of ISM absorption, \cite{Curtis} concluded in his analysis that none of the stars were MM candidates. 
\par
Thus, so far there has been no consensus on the discovery of a true MM star or even any very good candidate.
\par
\cite{Saar} also attempted to set down the criteria for classifying an MGM candidate; this analysis was inspired by \cite{Wright}'s question of whether \rphk\ was calibrated correctly since it did not account for the gravity and metallicity of the star. \cite{Saar} investigated the relationship between \rphk\ and metallicity, and they found that indeed the most inactive metal-rich stars had lower \rphk\ values than the most inactive metal-poor stars. \cite{Saar} set down new criteria for MM candidates: (1) The star is a bona-fide dwarf, (2) \rphk\ of the star should be lower than the  minimum seen for stars of similar metallicty and (3) HK variation as given by $\sigma/S_{\rm HK}\ \leq$ 2\% for a time period spanning $\geq$ 4 years of measurements. Thus, they implicitly assumed that MM events are rare and characterized by extraordinarily low \textit{R'}\textsubscript{HK} values.

\section{Data}
In order to compare results with the seminal Mt. Wilson survey, H\&K emission is usually expressed in the instrumental unit used by that survey, the S-Index, which essentially tracks the equivalent width of the emission in the H\&K line cores \citep{S-Index}. The strength of these Ca II emission lines is proportional to the heating of the Sun's chromosphere by the magnetic fields on the surface and has a direct correspondence with the strength of the magnetic activity and the area covered \citep{Leighton,Skumanich}. Another commonly used index is \rphk, which is the ratio of the chromospheric emission in H\&K lines to the total bolometric emission from the star so that it is independent of temperature and photospheric component of the star \citep{Noyes}.
\par 
For our analysis, we have used the HD 4915 spectra from 2006 until present, collected as a part of the California Planet Search Radial Velocity Survey \citep{CPS} using Keck/HIRES spectrograph \citep{HIRES}. The spectra are publicly available on Keck Observatory Archive (KOA).\footnote{\url{https://koa.ipac.caltech.edu}}\ The S-Index time series used have been derived from the spectra using methodology explained in \cite{Isaacson}.    

\begin{figure*}
	
   \plotone{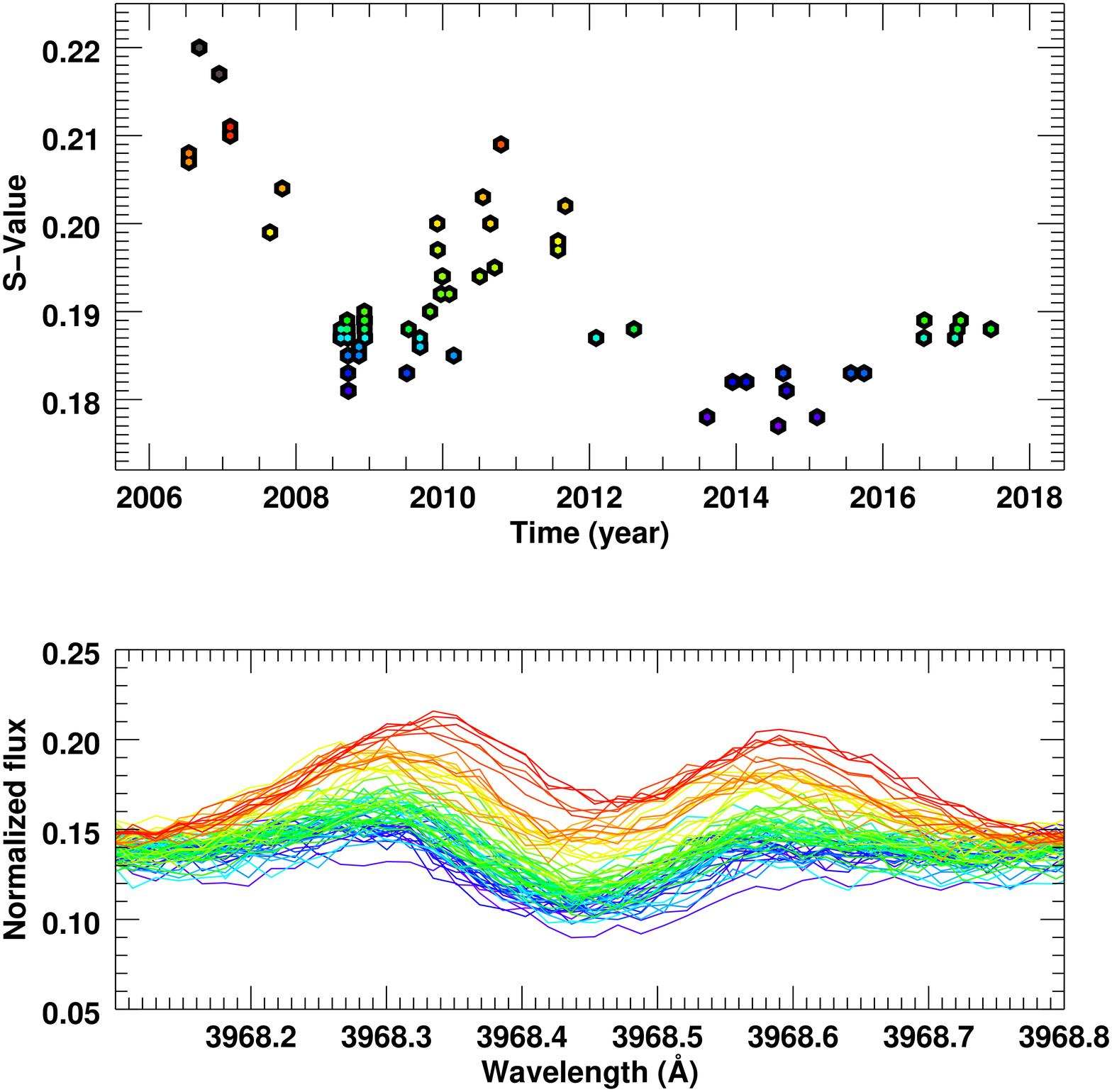}
       \caption{S-Indices of HD 4915 from 2006 until 2017 (top panel). The symbols are color coded according to their value. Y axis represents the S-Indices as calculated by \cite{Isaacson} with an uncertainty of $\pm\ 0.002$ and the X axis represents time in years. We see a trend of decreasing maxima as the magnetic activity of the star completes two cycles. We also note that the second minimum is lower than the first.
       The colors of the data points correspond to the respective colored spectrum in the spectral diagram of the star (bottom panel). The Y axis of the spectral diagram represents normalized flux and the X axis represents wavelength. The spectra is zoomed in to show Ca II H emission line at \textasciitilde 3968.4 {\AA} (an indicator of magnetic activity). The color coordination of the S-Indices with the spectra confirm the trend of decreasing amplitude of magnetic activity cycle.}
       \label{plot:spectra_sval}

\end{figure*}

Because the \cite{Butler} analysis of these same spectra find a slightly different S-Index time series that does not as obviously show evidence of entry into an MGM state, we have examined the spectra themselves to confirm that our S-Index measurements accurately reflect the emission in the H \& K lines.  The top panel of Fig. \ref{plot:spectra_sval} shows the S-Index time series color coded by value and the bottom panel shows the corresponding Ca II H line core spectra. We show this color coded correspondence to confirm a faithful tracking of the total emission in the line core by our S-Indices.

\section{HD 4915}
HD 4915 is a G5V star with an effective temperature of 5668 K, log g = 4.57 dex, vsini of 1.4 \kms, [M/H] of $-0.2$ dex and an isochrone derived age in the range of 0.6-4.4 Gyrs \citep{Specz}. The spectroscopic properties of HD 4915 differ from that of the Sun's by $-104$ K, 0.132 \citep{Sunspecz}, 0.2 km/s \citep{Sunvsini} and -0.2 dex, classifying it as a dwarf and also Sun-like.
\par
Over its last cycle, which we suspect to be its transition period into an MGM state, it has an average S-Index of 0.184 and and HK variation of 2.28\%. Following the analysis done by \cite{Noyes} for transforming S-Index to \rphk, and using $B-V = 0.66$ \citep{bmv}, we compute $\lrphk = -4.872$. 
\par
Using the activity--rotation--age relation formulated by \cite{Mamajek}, we determine the age of the star to be between 4.3 Gyrs (at the global minimum, $S = 0.176$) and 2.7 Gyrs (at the global maximum, $S = 0.22$), and to be 3.8 Gyrs over the last cycle. Thus, we note that it is difficult to attain an accurate age for a star that is entering an MGM state. Regardless, the range of ages is consistent with the isochrone age, confirming that the star is younger than the Sun.
\par
With \lrphk = $-4.877$ and [M/H] = $-0.2$, we note that it does not satisfy \cite{Saar}'s criteria of lying below the minimum \lrphk-[M/H] trend-line \textit{yet}. This could be explained \textit{if} HD 4915 is younger than the Sun, which would cause its average activity levels to be higher. On the other hand, it is a bona fide dwarf as classified by its temperature and gravity, and its HK variation of 2.28\% is close to the 2\% limit set down by \cite{Saar}.  
\par

\begin{figure}
\includegraphics[width=\linewidth]{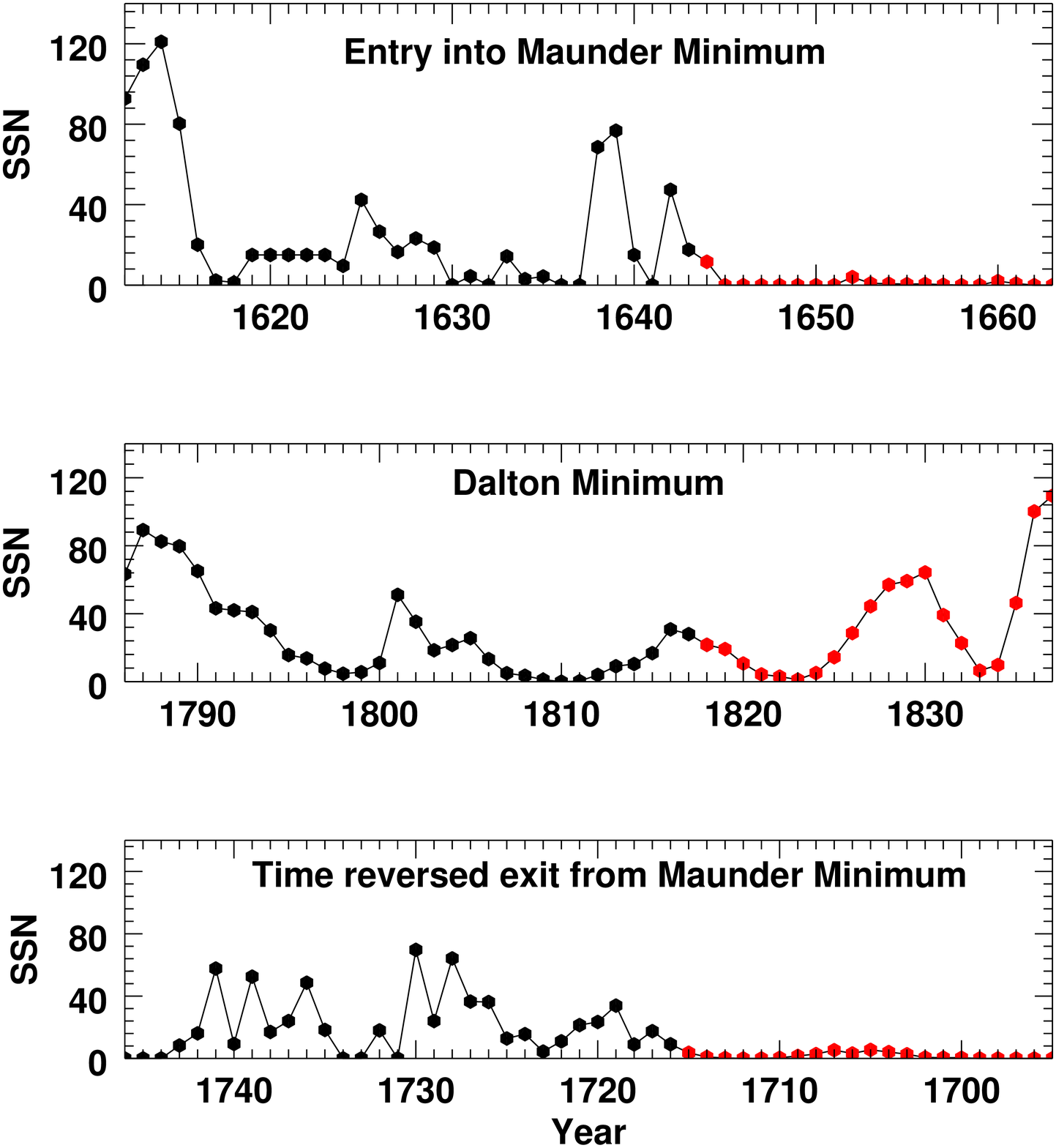}
\caption{Trend of Group Sunspot Number (GSN) time series during the Sun's two significant magnetic grand minima episodes since 1600 CE - Maunder Minimum and Dalton Minimum. The first and third panels show the GSN time series as the Sun enters and exits the Maunder Minimum, respectively. The second panel shows the GSN time series of the Sun during DM. The current trend of HD 4915 strongly resembles the trend of GSN as the Sun entered or exited the DM period, but if its low activity cycles continue for many more cycles or disappear entirely to be flat, it will be a relatively unambiguous Maunder minimum analog.}

\label{plot:sun}
\end{figure}

\section{A decreasing cycle strength reminiscent of solar MGM}
In this section, we analyze the S-Index time series of HD 4915 to categorize it as an MM candidate.  As the top panel of Figure~\ref{plot:spectra_sval} shows, we see a decrease in the amplitude of the activity cycle over the three apparent maxima until the amplitude is almost negligible. By examining the first 2 cycles and assume that the first measurements just capture the first cycle maximum, we estimate the period of the cycle to be 4 years and note the striking delay in the occurrence of the third maxima by almost 2 years. We also note that the second minimum appears to be significantly lower than the first minimum. These features of the S-Index time series strongly suggest a decreasing magnetic activity in the star. 
\par
To compare these features to that of the Sun, we look at the Sun's transitions into and out of MGM using its historical GSN record \citep[reconstructed by][]{Hoyt}. Figure \ref{plot:sun} shows entry into Maunder Minimum (first panel), the Dalton Minimum (middle panel) and exit from the Maunder Minimum (last panel). The black symbols represent the section of the GSN trend we are comparing with the past activity of HD 4915, and the red symbols represent the potential futures of the HD 4915 activity record if it were in an analogous state.
\par
The transition of the Sun into MM is characterized by a steep and sudden drop in GSN across two normal cycles (Figure \ref{plot:sun} top panel). GSN then remains close to 0 until towards the end of the MM at around 1700 CE, when it emerges from the Maunder Minimum over the course of two cycles and returns to its normal 11 year cycling state.  
\par
The S-Index time series of HD 4915 to date most resembles the transition of the Sun into the Dalton Minimum (DM) in Figure \ref{plot:sun} middle panel and marked by solid black dots. This transition is characterized by decreasing amplitude of maxima over three cycles and the third minimum occurring lower than the previous two (Fig. \ref{plot:spectra_sval}). 
\par
Finally, we compare the HD 4915 S-index time series to the exit of the Sun from the MM. This comparison is possible if we hypothesize that the entry and exit of the Sun's magnetic activity from a grand magnetic state are time symmetric. In the case of the Maunder Minimum event, the transition from MGM to a normal cycle as seen in Figure \ref{plot:sun} bottom panel is characterized by increasing amplitude of the magnetic activity (note the time reversed x-axis). This, too, bears some resemblance to the activity history of HD 4915.
\section{Discussion \& Conclusion}
The decline in the magnetic activity strength of HD 4915 is exciting because it renews the discussion of ways to identify MM candidates. As discussed in section 1.2, there have been numerous attempts over the last few decades to set down the criteria for an MGM candidate using S-Indices; however, these criteria have been of uncertain utility because we lack directly comparable data of the Sun's chromospheric activity during its MGM episodes.
\par
As indicated by \cite{Saar}, the surest way to characterize a star in MGM would be by observing it enter a long period (at least longer than its typical cycle period) of flat activity or low cycling activity and transition out of this period into a normal activity cycle. Alternatively, we could also make such a characterization by observing a star transition into a normal cyclic state from an anomalously low state. 
\par 
The discovery of a true Maunder-minimum analog would help us understand whether the state of the Sun's field during its MGM was a state of extraordinarily low global magnetic field strength, or just an extraordinarily weak and long series of magnetic maxima. Such an understanding would greatly inform searches for other Maunder minimum analogs.
\par
HD 4915 may, in fact, not be a star with a single, strong cycle period like the Sun. Indeed, we have inferred that it has a 4-year period under the assumption that our data capture the first partial cycle's maximum and what appears to be around 2 full "cycles" of varying morphology.  As has been shown by \cite{metcalfe}, many stars with cycle periods $<5$ years show stochastic or multi-period behavior that might be a better analogy to HD 4915 (see, for instance, HD 76151, Figs.~4.5 \& B.10 of \cite{Egeland} and HD 190406 Fig.~7 left column, 2nd from top; \citep{Hall2007}).  This underscores the need for future observations of this star to determine its magnetic behavior.
\par
We recommend immediate and future observations of chromospheric and coronal cooling indicators (such as the H \& K emission lines and X rays) for at least the next decade (two cycles of the star), especially since there is strong evidence that HD 4915 is entering a state of MGM right now. Further, it might be possible with precise photometry to estimate sunspot coverage for the star, which would help confirm the connection between sunspots and S-Indices for the star. {\it TESS} or other future space-based photometric missions might be able to determine if the star indeed has a pristine photosphere.
\par

\acknowledgments

We thank Eric Mamajek for useful discussions and for hunting down the early use of the term ``Dalton Minimum''.\footnote{\url{https://twitter.com/ericmamajek/status/929550539834007552}}  We also thank Thomas Beatty, Fabienne Bastien, Mark Giampapa, David Soderblom, and Larry Ramsey for useful discussions that improved the paper. 
 
The Center for Exoplanets and Habitable Worlds is supported by the
 Pennsylvania State University, the Eberly College of Science, and the
 Pennsylvania Space Grant Consortium.

The data presented herein were obtained at the W.M. Keck Observatory, which is operated as a scientific partnership among the California Institute of Technology, the University of California and the National Aeronautics and Space Administration. The Observatory was made possible by the generous financial support of the W.M. Keck Foundation.

This research has made use of NASA's Astrophysics Data System Bibliographic Services, the SIMBAD database, operated at CDS, Strasbourg, France, and the AstroBetter blog and wiki..

This material is based upon work supported by the NSF Postdoctoral Research Fellowship in Biology under Grant No.\ AST-1602662.

\facilities{Keck:I (HIRES)}

\bibliographystyle{yahapj}
\bibliography{references}

\end{document}